\documentclass[12pt,oneside]{article}
\usepackage{amsfonts,amssymb,graphicx}
\def\be{\begin{equation}}
\def\ee{\end{equation}}
\def\bea{\begin{eqnarray}}
\def\eea{\end{eqnarray}}
\def\<{\langle}
\def\>{\rangle}
\def\~{\tilde}
\def\s{\sigma}

\def\b{\beta}

\def\t{\tau}

\newcommand{\av}[1]{\mbox{{\it Av}}\left(#1\right)}

\begin{document}
\begin{center}
{\sc\Large replica equivalence in the edwards-anderson model}
\end{center}
\vspace{1cm}
\begin{center}{Pierluigi Contucci}\\ 
\vspace{.5cm}
{\small Dipartimento di Matematica} \\
    {\small Universit\`a di Bologna,
    40127 Bologna, Italy}\\
    {\small {e-mail: contucci@dm.unibo.it}}\\
\vspace{1cm}
Revised, July 24th, 2003 
\end{center}
%
%
\vspace{.5cm}
\begin{abstract}\noindent
After introducing and discussing the {\it link-overlap} between spin confi\-gurations
we show that the Edwards-Anderson model has a {\it replica-equivalent} quenched equilibrium 
state, a property introduced by Parisi in the description of the 
mean-field spin-glass phase which generalizes ultrametricity. Our method is based on the control of fluctuations 
through the property of stochastic stability and
works for all the finite-dimensional spin-glass models.
\end{abstract}
\newpage
In the description of the spin glass phase the standard quantity
usually investigated is the {\it overlap} between Ising spin
configurations:
\be
q(\s, \t) = \frac{1}{N}\sum_{i=1}^{N} \s_i\t_i \; .
\label{ov}
\ee
Within the mean-field approach such a quantity gives, in the so called
quenched ensemble, a complete description of the system and it is in fact
in terms of its distribution properties that the mean-field theory has
been developed and understood by Parisi \cite{MPV} and became successively 
accessible to a rigorous mathematical investigation starting from the seminal 
paper by Guerra \cite{G}.\\
The mean-field picture is described by two main features: first the overlap
quenched distribution has a support that includes the neighbor of zero
where the disorder is concentrated, second the distribution 
fulfills factorization-like properties \cite{MPV} and is, in particular,
completely  identified by the single overlap probability. The factorization properties have been
distinguished in two classes, {\it replica equivalent} and {\it ultrametric} (see \cite{P1}, \cite{P2} and
\cite{PRT}) the first being a subclass of the second in the sense that ultrametricity
implies replica-equivalence but, in general, not the viceversa. 
Replica equivalence is in fact introduced requiring to the overlap algebraic 
matrix ansatz $Q$ (defined in \cite{MPV}) to have any two rows (or columns) 
identical up to permutations. Such a condition is clearly satisfied by the 
replica symmetry breaking ansatz with its ultrametric structure but of course
it includes many non-ultrametric instances.

In \cite{AC}, \cite{GG} and \cite{C} it has been shown how
to derive replica-equivalence for mean-field spin-glass models from elementary thermodynamic 
properties like boundedness of fluctuations or investigating a new property of invariance
under random perturbations called stochastic stability. 
 
In this letter we show that the Edwards-Anderson spin-glass is 
{\it replica equivalent} with respect to the {\it bond-overlap} quenched distribution.
Our strategy is based on a reformulation of stochastic stability which holds for finite-dimensional systems.
Our method, which applies to every finite-dimensional Gaussian model, shows that 
the spin glass is fully described by the quenched distribution of a proper overlap and, 
with respect to it, is replica equivalent. The aim of the paper is then twofold: to establish new
features of the realistic spin-glass models and to stress 
the proper quantity to be investigated. 

To illustrate the physical meaning of our result we first derive the lowest order 
replica equivalence relation from the basic thermodynamic fact that the specific heat per 
particle is bounded everywhere (except at most on isolated singularities). Second we
show that the property of stochastic stability holds, when properly formulated, also 
in finite-dimensions and implies the entire set of replica equivalent identities
at every order.

For definiteness we consider the Edwards-Anderson spin-glass model but our
method is largely independent on the details of the interactions and at the end
we exhibit a wide class of finite-dimensional spin glass models to which 
our study apply {\it sic et simpliciter}. In the d-dimensional square lattice we 
study the the nearest-neighbors Hamiltonian
\be
H(J,\s) = -\sum_{(n,n')}J_{n,n'}\s_n\s_{n'} \; ,
\label{eag}
\ee
where the $J_{n,n'}$ represent the quenched disorder and are usually assumed to be
independent normal Gaussian variables. While the standard 
site-overlap between two spin configurations
$\s$ and $\t$ is the normalized sum of the local site-overlap (\ref{ov})
the bond-overlap is the normalized sum of the local nearest-neighbor overlap
\be
p(\s,\t) = \frac{1}{N_B}\sum_{(n,n')}\s_n\t_n\s_{n'}\t_{n'} \; ,
\label{bov}
\ee
$N_B$ being the number of nearest neighbor couples (bonds).
For a discussion of the relevance of bond-overlap (or link-overlap as it first appeared in \cite{MPRL1}) and for its use in
numerical experiments to study the low temperature phase of finite-dimensional Ising spin glasses one may see
\cite{MPRLZ} and \cite{MPRL2}. There is an obvious a priori advantage of the quantity (\ref{bov}) with
respect to (\ref{ov}): while a spin flip inside a bond-connected region changes $q(\s,\t)N$ of an amount proportional to
the region volume it only changes
$p(\s,\t)N_B$ of an amount proportional to the region surface. The previous observation fails of course
when the connectivity of the space grows with the volume like in the mean field cases.
In the Sherrington-Kirkpatrick model for instance the
usual overlap and the bond-overlap are related
by the algebraic formula $q^2=2p+1/N$ so that it is totally irrelevant which one of the two
is studied. On the contrary in finite-dimension it does not exist such a simple relation among the two
quantities even if from a bond configuration $\{\s_n\s_{n'}\}$ one may reconstruct the spin configuration $\{\s_n\}$
(up to a global sign) and vice-versa (see for this purpose the treatment of Gauge invariance in \cite{BF} and
\cite{N}). 

But definitely the deeper reason to introduce the bond overlap is related to the
mathematical properties of the Hamiltonian (\ref{eag}). Being a sum of Gaussian variables (the J's) it
is a Gaussian variable itself completely identified by 
its covariance matrix whose elements turn out to be proportional to the bond-overlap $p(\s,\t)$. 
Indicating by $Av$ the Gaussian average we have in fact:
\bea\nonumber
\av{H(J,\s) H(J,\t)} &=& \sum_{(n,n'), (m,m')}\av{J_{n,n'}J_{m,m'}}\s_n\s_{n'}\t_m\t_{m'}  \\ \nonumber
&=& \sum_{(n,n'), (m,m')}\delta_{(n,n')(m,m')}\s_n\s_{n'}\t_m\t_{m'} \\
&=& N_B \, p(\s,\t) \; .
\label{cova}
\eea
More specifically we will work with the Hamiltionian (\ref{eag}) as with a family of $2^N$ Gaussian
variables $\{H_\s\}$ (one for each configuration $\s$) whose joint distribution is specified
by the $2^N\times 2^N$ square matrix of elements $p(\s,\t)$ (for this perspective in the mean-field case see \cite{CDGG}).

The previous observation says that all the typical quantities that are 
derived from the free energy like the internal energy the specific heat etc.
are described by the bond-overlap moments with respect to the quenched
measure. The same situation occurs in parallel for the mean-field case with respect to
its own covariance matrix which is the square power of the standard site-overlap. Let for completeness show how in the
Edwards-Anderson case the internal energy and specific heat are related to the quenched average of the matrix $p$. 
As the computation
is going to illustrate our result does not depend on the detailed structure of the Hamiltonian
as far as it is Gaussian.
The quenched internal energy
\be
U(\beta) \; = \; \av{\frac{\sum_\s H_\s e^{-\b H_\s}}{\sum_{\s'} e^{-\b H_{\s'}}}} \; 
\label{ie}
\ee
can be related to the bond-overlap moment using the elementary rule of integration by parts for correlated Gaussian
variables $\{\xi_i\}$ with covariances $c_{i,j}$ which states that for every bounded function $f$
\be
\av{\xi_i\cdot f}\,=\,\av{\sum_{j} c_{i,j}\cdot\frac{\partial
f}{\partial \xi_j}} \; .
\label{iipp}
\ee 
Applying the (\ref{iipp}) to the right hand side of (\ref{ie}) gives
\be
\av{\frac{H_\s e^{-\b H_\s}}{\sum_{\s'} e^{-\b H_{\s'}}}} \; = \; N_B \av{\sum_\t p(\s,\t)\frac{\partial}{\partial H_\t}\frac{e^{-\b
H_\s}}{\sum_{\s'} e^{-\b H_{\s'}}}} \; ,
\label{ip}
\ee
and after the straightforward computation of the derivative we obtain
\be
\frac{U(\beta)}{N_B} \; = \; -\b\left(1- \av{\frac{\sum_{\s,\t}p(\s,\t)e^{-\b (H_\s + H_\t)}}{\sum_{\s,\t}e^{-\b (H_\s +
H_\t)}}}\right) \; .
\label{ieod}
\ee
The (\ref{ieod}) shows that the internal energy can be computed
by first averaging the $p(\s,\t)$ with respect to the random Gibbs-Boltzmann state
over {\it two copies} of the system and then quenching the disorder by the Gaussian average.
The final resulting operation is a probability measure (the so called quenched state $E$)
over the matrix element $p(\s,\t)$
\be
E(p_{1,2}) \; = \; \av{\frac{\sum_{\s,\t}p(\s,\t)e^{-\b (H_\s + H_\t)}}{\sum_{\s,\t}e^{-\b (H_\s +
H_\t)}}} \; ,
\label{azze}
\ee
\be
\frac{U(\beta)}{N_B} \; = \; -\b\left(1- E(p_{1,2})\right) \; .
\label{ieod2}
\ee
More generally for higher order quantities like the specific heat etc. one introduces
the quenched measure over an arbitrary number $r$ of copies. For instance the computation
of the specific heat is related, among others, to the moment
\be
E(p_{1,2}p_{2,3}) \, = \, 
\av{
\frac{\sum_{\s_1,\s_2,\s_3}p(\s_1,\s_2)p(\s_2,\s_3)e^{-\b (H_{\s_1}+H_{\s_2}+H_{\s_3}) }}{\sum_{\s_1,\s_2,\s_3}e^{-\b
(H_{\s_1}+H_{\s_2}+H_{\s_3})}}} \; .
\label{czz}
\ee
From the definition of the specific heat per particle and using again just the rule of 
integration by parts (\ref{iipp}) we get from (\ref{ieod2})
\begin{eqnarray}
c(\beta) &=& \frac{d}{d\beta}\frac{U(\beta)}{N_B} = -\, (1-E(p_{1,2})) \, + \,
\b\frac{d}{d\beta}E(p_{1,2})
\, = \\
\nonumber
\, &=& -\,  (1-E(p_{1,2}))   + 2\beta N_B E(p_{1,2}^2-4p_{1,2}p_{2,3}+3p_{1,2}p_{3,4}) \; .
\end{eqnarray}
Due to the convexity of the free energy the specific heat per particle is a bounded quantity (a part at most on isolated
singularities \cite{Ru,G}) and the (\ref{czz}) shows that in the thermodynamic limit
($N_B\to \infty$) the quenched state has to fulfill the identity
\be
E(p_{1,2}^2-4p_{1,2}p_{2,3}+3p_{1,2}p_{3,4}) = 0 \; ,
\label{id1}
\ee
with a rate of decrease of at least $N^{-1}$.
The previous relation is the lowest order replica equivalence identity (see \cite{P1} and \cite{P2}).
Before introducing a general criterion which reproduce the whole set of those identities
we want to stress that the previous discussion shows that the $r\times r$ matrix $P$ of elements
\be
p_{l,m} \, = \, p(\sigma^{(l)}, \sigma^{(m)})  \
\ee
together with its probability measure $E$ fully describes the Edwards-Anderson model
in the sense that the moments of $P$ like $E(p_{1,2})$, $E(p^2_{1,2}p_{1,3})$ etc. represent the 
entire set of physical observables of the theory.

Let now develop an approach to stochastic stability for finite dimensional systems which runs
parallel to the one introduced in \cite{AC} for the mean field case. 
The starting point is the observation that the addition to the
Hamiltonian of an independent Gaussian term of finite size:
\be
h(\tilde{J},\s) = \frac{1}{\sqrt{N_B}}\sum_{(n,n')}\tilde{J}_{n,n'}\s_n\s_{n'}
\label{fsh}
\ee
amounts to a slight change in the temperature. In fact for the sum law of independent Gaussian 
variables one has that in distribution it holds the relation
\be
\beta H(J,\s) + {\lambda} h(\tilde{J},\s) \ 
=
\beta'(\beta,\lambda) H(J',\s) \; ,
\label{eq:delta}
\ee
with
\be
\beta'(\beta,\lambda) \, = \, \sqrt{\beta^2 + \frac{\lambda^2}{N_B}} \; ,
\ee
so that, indicating by $E_\lambda^{(\beta)}$ the quenched state of Hamiltonian $\beta H(J,\s) + \lambda
h(\tilde{J},\s)$ the (\ref{eq:delta}) implies
\be
E_{\lambda}^{(\beta)} \, = \, E^{(\beta')}
\; \; .
\label{propri}
\ee
Taking the thermodynamic limit (see \cite{CG}) of the (\ref{propri}) 
and observing that
\be
\lim_{N_B\to\infty}\b'=\b
\ee
we obtain
\be
E_{\lambda}^{(\beta)} \, = \, E^{(\beta)}
\; \; ,
\label{ssea}
\ee
for all values of $\b$ a part, at most, isolated singularities.
Such a property, introduced in the mean field case in \cite{AC}, is called {\it stochastic stability} \cite{P3} 
and was later investigated in \cite{FMPP1} and \cite{FMPP2} to determine a relation between the off-equilibrium
dynamics and the static properties. Stochastic stability has important consequences for the quenched state. In particular it says
that the second derivative of the $\lambda$-deformed moments have to be zero (the first derivative being zero for
antisymmetry). Let apply it for instance to the moment $E(p_{1,2})$. The elementary computation which uses only the Wick
rule gives
\be 
\frac{d^2}{d\lambda^2} E_{\lambda }(p_{1,2})|_{\lambda=0} =
2 E(p_{1,2}^2-4p_{1,2}p_{2,3}+3p_{1,2}p_{3,4}) \; ,
\label{ppac}
\ee
so that from (\ref{ssea}) we have (in the infinite volume limit)
\be 
E(p_{1,2}^2-4p_{1,2}p_{2,3}+3p_{1,2}p_{3,4}) \, = \, 0\; .
\label{ppac2}
\ee
Applying analogously stochastic stability to the moment $E(p_{1,2}p_{2,3})$ we obtain
\be 
E(2p_{1,2}^2p_{2,3}+p_{1,2}p_{2,3}p_{3,1}-6p_{1,2}p_{2,3}p_{3,4}+6p_{1,2}p_{2,3}p_{4,5} -3p_{1,2}p_{2,3}p_{2,4}) \,
=
\, 0\; .
\label{ppac3}
\ee
The (\ref{ppac2}) and (\ref{ppac3}) are two of the possible identities found in the framework
of replica-equivalence \cite{P1}. The application of
stochastic stability to the whole set of moments produces the whole set of replica equivalence identities (this
is a purely combinatorial argument and it may be seen in \cite{C}).

We want to observe that our result is different from the one mentioned 
in \cite{G}. The author there (at the end of section 4) suggests a method to prove some factorization
identities for the standard overlap (\ref{ov}) for a general spin model. The result is achieved with the addition
to the Hamiltonian of a term
proportional to a mean-field spin-glass interaction and sending the interaction strength to zero after the thermodynamic
limit is considered. The resulting state turns out to be replica-equivalent for the standard site-overlap 
but the addition of a mean
field perturbation could, in principle, select a sub-phase of the whole equilibrium state. Our result instead
obtains replica equivalence with respect to the bond-overlap for the whole quenched state.

Our scheme allows the treatment of the spin-glass Hamiltonians which include many-body interactions as well as unbounded
spins variables:
\be
H(J,\s) = - \sum_{X}J_X\s_X \; ,
\label{sgp}
\ee
where 
\be
\s_X=\prod_{i\, \in X}\s_i \; ,
\ee
and the $J_X$'s are independent Gaussian variables with zero mean and variance $\Delta^2_X$.
A simple calculation like the one in (\ref{cova}) gives
\bea\nonumber
Av(H(J,\s) H(J,\t)) &=& \sum_{X, Y}Av(J_{X}J_{Y})\s_X\t_Y \\ 
&=& \sum_{X}\Delta_X^2\s_X\t_X \; .
\label{covag}
\eea
The replica equivalence identities that have been derived in the mean field case \cite{AC} in terms of the standard
site-overlap and that we obtained here in terms of the bond-overlap for the Edwards-Anderson model can be proved
for the model of Hamiltonian (\ref{sgp}) in terms of the generalized multi-overlap:
\be
{\widetilde P}(\s,\t) \, = \, \frac{1}{{\cal N}}\sum_{X}\Delta_X^2\s_X\t_X \; ,
\ee
where we have indicated by ${\cal N}$ the number of interacting subsets.

Summarizing we have shown that replica-equivalence holds in the Edwards-Anderson model
as well as in all Gaussian finite-dimensional spin-glass models when properly formulated in terms
of the relative overlap. As the physical intuition
suggests the result is robust enough to remain true also when the disorder $J$ is chosen at random
from a $\pm 1$ Bernoulli disorder or for more general centered distributions (see \cite{CG}). 
Our result holds at every temperature a part at most at isolated singularities and, in particular,
it cannot make predictions at exactly $T=0$. We want to remark moreover that replica equivalence
does not identify uniquely the low temperature phase. It is in fact compatible with both the so called
droplet picture (see \cite{FH}, \cite{NS}) and with the replica symmetry breaking one \cite{MPV}.
Nevertheless it provides a rigorous proof of an infinite family of identities among overlap moments 
and, especially, it clearly points out the role and the importance of the suitable overlap to describe the
model with. 
\\\\ {\sc acknowledgments}\\
The author thanks S.Graffi, F.Guerra, E. Marinari, H.Nishimori and G.Parisi for interesting discussions and the Tokyo Institute
of Technology for the kind hospitality.

\end{document}